\documentclass[aps,prb,twocolumn,showpacs]{revtex4}

\usepackage{graphicx}

\begin{document}

\title{\bf Decoherence in Josephson vortex quantum bits}
\author{Ju H. Kim and Ramesh P. Dhungana}
\affiliation{Department of Physics, University of North Dakota,
Grand Forks, ND 58202-7129}

\author{Kee-Su Park}
\affiliation{National Creative Research Initiative Center for
Superfunctional Materials, Chemistry Department\\
Division of Molecular and Life sciences, Pohang University of
Science and Technology, Pohang 790-784, Korea }

\begin{abstract}
We investigated decoherence of a Josephson vortex quantum bit
(qubit) in dissipative and noisy environment. As the Josephson
vortex qubit (JVQ) is fabricated by using a long Josephson junction
(LJJ), we use the perturbed sine-Gordon equation to describe the
phase dynamics representing a two-state system and estimate the
effects of quasiparticle dissipation and weakly fluctuating critical
and bias currents on the relaxation time $T_1$ and on the dephasing
time $T_\phi$.  We show that the critical current fluctuation does
not contribute to dephasing of the qubit in the lowest order
approximation.  Modeling the weak current variation from magnetic
field fluctuations in the LJJ by using the Gaussian colored noise
with long correlation time, we show that the coherence time $T_2$ is
limited by the low frequency current noise at very low temperatures.
Also, we show that an ultra-long coherence time may be obtained from
the JVQ by using experimentally accessible value of physical
parameters.
\end{abstract}


\maketitle

\section{Introduction}

Novel superconducting quantum bits (qubits), such as charge (i.e.,
Cooper-pair box),\cite{charge} flux,\cite{flux} and phase
qubits,\cite{phase} are good candidates for quantum information
processing because these can be manufactured, controlled, and
scaled more easily.  Tens of quantum oscillations had been
observed in these qubits, but a low level of decay which yields
thousands of coherence oscillations is essential for realization
of quantum computation.  The longest coherence time of 0.5 $\mu$s
has been reported for the quantronium\cite{chfl} (i.e., hybrid
charge-flux qubit), but longer time may still be necessary. As
many decoherence sources reduce the quantum oscillations, the
measured value of the coherence time for the superconducting
qubits is substantially shorter\cite{Har} than that predicted by
the simplest models of decoherence and that needed for the
operation of a quantum computer.  This requirement of obtaining
ultra-long coherence time in the presence of the interaction
between the qubit system and noisy environment is one of major
challenges.

Understanding the mechanisms of decoherence became a focus of much
attention recently, and it remains an important challenge for the
superconducting qubits.  An ideal solution is to isolate the qubit
system from uncontrolled degree of freedom in its environment and
in the device itself.  However, it is difficult to isolate the
qubit system completely from the decoherence sources.  These
sources include, but may not be limited to, background charge
fluctuations in the substrate,\cite{shot} fluctuations in the
tunnel barrier which produce microscopic tunneling
resonance,\cite{micro} and fluctuating electromagnetic background.
Also, low frequency variation in the critical current\cite{Har} is
present in all superconducting qubits.  One way to obtain the
ultra-long coherence time is to use the Josephson vortex qubit
(JVQ) since it msy be immune to these sources.

Recently, JVQ has been proposed as a new superconducting
qubit.\cite{Cl}  This qubit has two important advantages over
other superconducting qubits.  First, coupling between the qubit
system and the decoherence sources is weak at very low
temperatures.  For example, a Josephson vortex (i.e., fluxon) in a
uniform long Josephson junction (LJJ) does not generate any
radiation during its motion and is almost decoupled from other
electromagnetic excitations in the junction.  Also the qubit is
immune to fluctuations in the critical current. Consequently,
quantum coherence can be maintained much longer than other qubits
which are susceptible to these decoherence sources. Second, as the
fluxon dynamics in LJJ is described by using the perturbed
sine-Gordon equation,\cite{MS} the decoherence sources for the
qubit may be easily identified. For example, two important sources
are the quasiparticle dissipation and weak current noise. Hence,
the coherence time may be estimated more easily, but the effect of
these sources has not yet been estimated for the JVQ.

\begin{figure}[t]
\includegraphics[width=6.5cm]{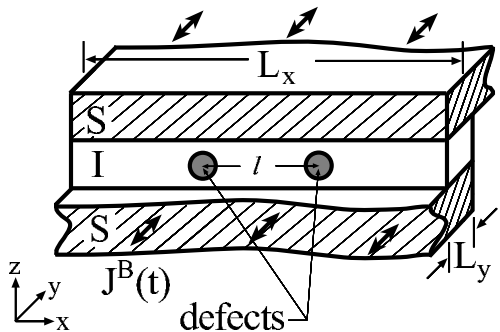}
\caption{ A LJJ stack is shown schematically as alternating layers
of superconductors ($S$) and insulator ($I$).  $L_x$ and $L_y$
denote the dimensions in $x-$ and $y-$direction, respectively.
$J^B(t)$ denotes the time dependent bias current density. The
filled dashed circles represent the microresistors which behave as
pinning centers for fluxon.} \label{fig1}
\end{figure}

The JVQ exploits the property of fluxon,\cite{KI,Sh} which behaves
as a quantum particle at ultra-low temperatures. The fluxon
trapped in a controllable potential well in a single annular LJJ
shows (i) energy quantization in the potential well and (ii)
macroscopic quantum tunneling (MQT) from a metastable
state.\cite{Wa}  Also, it was shown that the two quantum states
for the qubit can be created by using a heart-shaped\cite{Wa}
annular LJJ and by trapping a fluxon in a magnetic
field-controlled double-well potential.  These quantum states may
also be created using a linear LJJ with two closely implanted
defect sites in the insulator layer.\cite{KI}  A Nb-AlO$_x$-Nb
junction may be used to fabricate the JVQ, as shown schematically
in Fig. 1.  The dimensions of the junction, compared to the
Josephson length $\lambda_J$, are $L_x \gg \lambda_J$ and $L_y \ll
\lambda_J$.  The separation distance $\ell$ between the defect
sites must be larger than the critical distance $\ell_o$, as
discussed below.  Also, the preparation of initial state and the
read-out of the final qubit state may be performed by using
classical circuits.\cite{KU}

As the effect of the decoherence sources in the LJJ is described
by using the perturbed sine-Gordon equation, the coherence time
$T_2$,
\begin{equation}
{1 \over T_2}={1 \over 2T_1} + {1 \over T_\phi}~,
\end{equation}
for the JVQ may be estimated without making further assumptions
about the nature of the qubit-environment interaction. Here $T_1$
is the relaxation time, and $T_\phi$ is the dephasing time.  Since
$T_\phi$ is most sensitive to the decoherence sources, extending
$T_\phi$ for the superconducting qubits is important for quantum
computing applications.  We compute $T_2$ by accounting for the
two sources: (i) quasiparticle dissipation and (ii) weak current
noise.  It is noted that the JVQ may also couple to other sources,
such as microwave and phonon radiation, but the effect of these
sources is expected to be small as they lead to higher
order\cite{Io} contribution in the perturbation expansion than
that considered in the present paper. In the JVQ, the weak current
noise represents the low frequency magnetic field and current
fluctuations in the junction.  We focus on slow fluctuations since
the qubits suffer from the presence of strong low frequency noise
sources. The effects of these two decoherence sources have been
investigated also for other superconducting qubits\cite{Xu} and
were found to be important.

In this paper, {\it we show that the JVQ can yield ultra-long
coherence time because it couples very weakly to noisy environment
at low temperatures.}  Before proceeding further, we outline the
main results. i) Starting from the perturbed sine-Gordon equation,
we show that {\it the critical current fluctuation does not couple
to the JVQ within the lowest order approximation. Consequently, this
fluctuation effect does not lead to decoherence of the qubit.} ii)
We show that {\it $T_2$ at ultra-low temperatures is determined by
the low frequency current noise since the dissipation effect due to
qubit-environment coupling is exponentially small.} iii) Using
experimentally obtained physical parameters, we show that {\it the
effect of this current noise on decoherence is weak in the JVQ.}
This weak coupling between the JVQ and current noise leads to {\it
the coherence time of several tens of microseconds for the JVQ.}

We outline the remainder of the paper.  In Sec. II, we express the
phase dynamics of LJJ in the collective coordinate representation
and transform the perturbed sine-Gordon equation onto the
double-well potential problem.  In Sec. III, we obtain the
two-state system, described by the spin-boson model with low
frequency current noise.  Here, the quasiparticle dissipation is
described by using Ohmic environment, and the effects of low
frequency noise in LJJ is described by using the fluctuating weak
bias current. In Sec. IV, we derive the relaxation time ($T_1$)
and the dephasing time ($T_\phi$) due to these two decoherence
sources. In Sec. V, we estimate numerically the coherence time
($T_2$) by using the experimentally accessible parameters for LJJ
and compare it to that obtained for other superconducting qubits.
Finally, in Sec. VI, we summarize the result and conclude.

\section{long Josephson junction qubit}

In this section, we discuss how the LJJ may be used to obtain a
JVQ by starting from the perturbed sine-Gordon equation.  First, a
double-well potential for the fluxon needs to be created in the
LJJ to obtain the two quantum states of the JVQ.  Several
approaches are used to accomplish this.  Each of these approaches
yields a slightly different form of the potential function.  In
this paper, we consider the approach of implanting two closely
spaced microresistors in the insulator ($I$) layer. When the
fluxon does not have enough kinetic energy, the microresistor
attracts the fluxon and traps it at the defect site. The effects
of quasiparticle dissipation and low frequency weak current noise
may be examined by starting with
\begin{equation}
{\partial^2 \varphi \over \partial x^2} - {\partial^2 \varphi
\over \partial t^2} - \sin \varphi = {\cal F}~,
\label{sGlay1}
\end{equation}
where $x$ and $t$ are the dimensionless coordinates in units of
$\lambda_J$ and $\omega_p^{-1}$, respectively.  Here $\omega_p$
denotes the plasma frequency.  The dynamic variable $\varphi$
represents the difference between the phase of order parameter for
the superconductor ($S$) layers.  The perturbation term ${\cal
F}=\beta (\partial \varphi/\partial t)
-\beta_s(\partial^3\varphi/\partial t \partial^2x) + f(t) + {\bar
{\delta J_c}} (t) \sin\varphi -\sum_i \epsilon_i\delta(x-x^o_i)
\sin\varphi$ includes the effects due to quasiparticle dissipation
($\beta$ and $\beta_s$), bias current ($f=J^B/J_c$), critical
current fluctuation ($\delta {\bar J}_c(t)=\delta J_c(t)/J_c$),
and microresistors ($\epsilon_i= (J_c-J'_c)l_b/J_c\lambda_J$).  We
note that the critical current $J_c(t)$ may be expressed as the
sum of uniform ($J_c$) and weak fluctuation parts ($\delta
J_c(t)$): $J_c(t)=J_c+\delta J_c(t)$.  Here $x^o_i$, $J^B$,
$J'_c$, and $l_b$ ($\ll \lambda_J$) denote, respectively, the
position of inhomogeneity in the $I$ layer, the bias current
density, the modified critical current density at the defect site,
and the length of the LJJ in which $J_c$ is modified. In the
discussion below, we neglect the term $\beta_s (\partial^3 \varphi
/\partial t \partial^2 x)$ due to the quasiparticle (surface)
current along the junction layer, for simplicity, since both
$\beta (\partial\varphi/\partial t)$ and $\beta_s (\partial^3
\varphi /\partial t \partial^2 x)$ terms yield similar dissipation
effects.

The perturbation term $\cal F$ in Eq. (\ref{sGlay1}) is small, and
consequently, it does not change the form of the kink within the
framework of the lowest approximation.\cite{MS}  We describe the
motion of the fluxon in terms of the center coordinates $q(t)$,
which are obtained by neglecting $\cal F$.  In the absence of the
perturbation terms (${\cal F}=0$), the fluxon solution to Eq.
(\ref{sGlay1}) in the non-relativistic limit (i.e., $v \ll 1$) is
given by
\begin{equation}
\varphi (x,t) \approx 4 \tan^{-1}\left[ e^{\gamma (v) [x-q(t)]}
\right]~,
\label{soliton}
\end{equation}
where $\gamma^{-1}(v)= \sqrt{1-v^2}$, $q(t)=v t$ denotes the
center coordinate of the fluxon, and $v$ is the fluxon speed in
units of Swihart velocity $c$.  $q(t)$ is also known as collective
coordinate and represents a dynamical variable. We note that this
solution represents propagation of nonlinear wave as a ballistic
particle and that the perturbation terms in ${\cal F}$ only affect
dynamics of the center coordinates.

Applying the kink solution of Eq. (\ref{soliton}) to Eq.
(\ref{sGlay1}) for carrying out the classical perturbation theory
within the framework of lowest order approximation,\cite{MS} we
obtain the equation of motion for the center coordinate $q(t)$ in
the nonrelativistic limit as
\begin{equation}
M {d^2q(t) \over dt^2} + \beta M {dq(t) \over dt} + {\partial V(q)
 \over \partial q} = 0~.
 \label{classic}
\end{equation}
Here $M=8$ denotes the rest mass of the fluxon which is obtained
by inserting the waveform of Eq. (\ref{soliton}) into the
Hamiltonian corresponding to the unperturbed sine-Gordon equation
(i.e., Eq. (\ref{sGlay1}) with ${\cal F}=0$).\cite{MS}  $V(q)$ is
the effective potential for the fluxon due to the non-dissipative
perturbation terms in ${\cal F}$.

The phase dynamics in the center coordinate may be seen easily
from the Euclidean Lagrangian, ${\cal L}= {\cal L}_o+{\cal L}_P$
where ${\cal L}_o$ and ${\cal L}_P$ describes the unperturbed
phase dynamic of LJJ and the perturbation contribution,
respectively. The unperturbed part of the Lagrangian ${\cal L}_o$
is given by ${\cal L}_o=\int (dx/2) [(\partial \varphi /\partial
\tau)^2 + (\partial\varphi /\partial x )^2 + 2(1-\cos\varphi)]$.
The perturbation part of the Lagrangian ${\cal L}_P={\cal
L}_{nd}+{\cal L}_{d}$ can be expressed as the sum of the
non-dissipative part (${\cal L}_{nd}$) due to the bias current,
critical current fluctuation and inhomogeneities, and the
quasiparticle dissipation part (${\cal L}_{d}$). The
non-dissipative contribution may be expressed as ${\cal
L}_{nd}={\cal L}_{bias}+{\cal L}_{\delta J_c} +{\cal L}_{pin}$.
Here ${\cal L}_{bias}=\int dx f \varphi$, ${\cal L}_{\delta J_c} =
{\bar{\delta J_c}}(\tau) \int dx (1-\cos \varphi )$, and
\begin{equation}
{\cal L}_{pin}=\sum_{i=1}^2 \int dx~ \epsilon_i
\delta(x-x_i^o)(1-\cos\varphi)
\end{equation}
are the Lagrangian for the bias current, critical current
fluctuation, and two defect sites, respectively.  Following
Caldeira and Leggett,\cite{CL} we account for the quasiparticle
dissipation (i.e., $\beta$) by representing the environment as a
heat bath.   The heat bath may be described as a reservoir of
harmonic oscillators with generalized momenta $P_i$ and
coordinates $Q_i$. The dissipation effect due to coupling between
the phase variables $\varphi$ and the heat bath is described as
\begin{equation}
{\cal L}_{d}=\int dx \sum_{i} \left[ {P_i^2 \over 2m_i} +
{m_i\omega_i^2 \over 2} \left( Q_i - {c_i \varphi \over m_i
\omega_i^2} \right)^2 \right]~.
\end{equation}
Here, the heat bath parameters $m_i$, $\omega_i$ and $c_i$
characterize the reservoir's spectral function $J_\beta(\omega)$,
which is written as
\begin{equation}
J_\beta(\omega) = {\pi \over 2} \sum_i {c_i^2 \over m_i \omega_i}
~ \delta(\omega - \omega_i) = \beta\omega~. \label{Spec}
\end{equation}
This spectral function reproduces the dissipation term in Eq.
(\ref{classic}) when the heat bath degrees of freedom are
integrated out.

We describe the fluxon dynamics by using semiclassical theory as
usually done\cite{GJS} and by reexpressing the partition function,
${\cal Z}=\int {\cal D}[\varphi] \exp\{-S[\varphi]\}$, with
$S[\varphi]=\int d\tau {\cal L}$, in terms of the collective
coordinates $q(t)$ as ${\cal Z}=\int {\cal D} [q] \exp\{-S[q]\}$.
We take the perturbation expansion in terms of $\beta$, $f$, $\bar
{\delta J_c}$ and $\epsilon_i$, assuming that all of these
parameters are small. The lowest order contribution from this
expansion is obtained by substituting the soliton solution of Eq.
(\ref{soliton}) to the action $S[\varphi]$ since the perturbation
term $\cal F$ does not modify the soliton waveform in the lowest
order.\cite{MS,GJS} In this center coordinate representation, the
bias current contribution to the action (i.e.,
$S_{bias}[\varphi]=\int d\tau {\cal L}_{bias}$) yields\cite{KI}
\begin{equation}
S_{bias} [q]=S_{bias}[\varphi (x-q)] - S_{bias} [\varphi (x)]=
-\int d\tau(2\pi fq)~.
\end{equation}
Here the constant $S_{bias} [\varphi (x)]$ is subtracted since we
have chosen the origin of potential energy for the center
coordinate at $q=0$.  On the other hand, the critical current
fluctuation contribution (i.e., $S_{\delta J_c}=\int d\tau {\cal
L}_{\delta J_c}$) becomes
\begin{equation}
S_{\delta J_c} [q]=S_{\delta J_c}[\varphi (x-q)]= S_{\delta J_c}
[\varphi (x)]~,
\end{equation}
indicating that $S_{\delta J_c} [q]$ is independent of $q$.  This
suggests that the critical current fluctuations do not modify the
fluxon potential.  {\it Within the lowest order approximation, the
low frequency noise corresponding to the critical current
fluctuation does not couple to the JVQ, and it does not contribute
to decoherence}.  Also, other perturbation contributions
$S_{pin}=\int d\tau {\cal L}_{pin}$ and $S_d=\int d\tau {\cal
L}_d$ can be expressed in the $q$ representation. Combining these
perturbation contributions, we may express the partition function
as ${\cal Z}=\int {\cal D}[q(\tau)]\exp \{-S_{eff}[q(\tau)]\}$,
where the effective action $S_{eff} [q]$ is given by
\begin{eqnarray}
S_{eff}[q] = \int d\tau \left[ {1 \over 2} M {\dot q}^2 + V(q)
\right] ~~~~~
\nonumber \\
+ {M \over 2} \int d\tau \int d\tau' {\cal K}(\tau-\tau') [
q(\tau) -q(\tau') ]^2~.
\label{action}
\end{eqnarray}
The quasiparticle dissipation effect (i.e., $\beta$) at the finite
temperature $T$ is described by the kernel ${\cal K}(\tau)$
\begin{equation}
{\cal K}(\tau)={1\over \pi} \int_0^\infty d\omega~J_\beta
(\omega)~{\cosh(\omega/2T - \omega \vert \tau \vert ) \over
\sinh(\omega/2T)} ~.
\end{equation}
Here  we set $\hbar=k_B=c=1$ for convenience.  The potential
function $V(q)$ for the fluxon in the collective coordinates is
given by
\begin{equation}
V(q)= -2\pi f(t) q - {2 \epsilon_1 \over \cosh^2 \left(q-{\ell
\over 2}\right)}- {2\epsilon_2 \over \cosh^2 \left(q+{\ell \over
2}\right)}~,
\label{poten}
\end{equation}
where $\ell$ is the separation distance between two defect sites,
as shown in Fig. 1. The fluxon potential $V(q)$ includes the
potential tilting effect of the bias current ($f$) and the pinning
effect ($\epsilon_i$) of the two defect sites.

\begin{figure}[t]
\includegraphics[width=6.5cm]{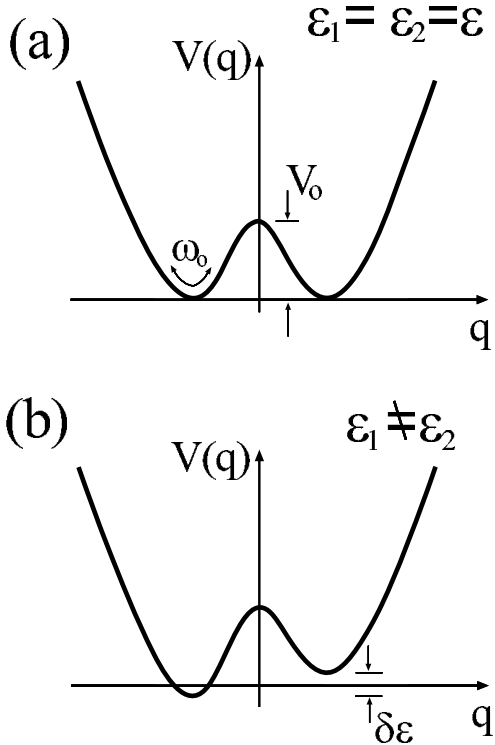}
\caption{ A double-well potential due to two microresistors is
schematically shown for (a) $\epsilon_1=\epsilon_2=\epsilon$
(symmetric) and (b) $\epsilon_1 \ne \epsilon_2$ (asymmetric). The
potential barrier and the oscillation frequency at the stable
minimum are denoted by $V_o$ and $\omega_o$, respectively. Here
$f(t)$ is set to zero.} \label{fig2}
\end{figure}

The potential function $V(q)$ of Eq. (\ref{poten}) for $f(t)=0$
has two noteworthy features: (i) finite number of bound states,
and (ii) double-well structure.  For physical values of
$\epsilon_i$, at most, several states may be trapped by the fluxon
potential.  This can be seen easily from the energy eigenstate of
the trapped fluxon via a single microresistor,\cite{LL} which is
given by
\begin{equation}
{\cal E}_n = -{1 \over 64} \left[ -(1+2n)+ \sqrt{1+128\epsilon}~
\right]^2
\end{equation}
where $n=0, 1, 2, \cdot\cdot\cdot$.  For $\epsilon=0.27$, only
$n=$0, 1, and 2 states, corresponding to the eigenstate energy
${\cal E}_0=$-0.383, ${\cal E}_1=$-0.136, and ${\cal
E}_2=$-0.0142, respectively, are bounded by the potential. Also,
the double-well structure can be seen easily by setting
$\epsilon_1 = \epsilon_2 =\epsilon$ (i.e., symmetric double-well)
(see Fig. 2(a)) and by expanding the function $V(q)$ about the
critical separation distance $\ell_o$.  We note that a small
asymmetry (or bias) of ${\bar \epsilon} \approx 8
q_o\delta\epsilon$ may be easily introduced, as shown in Fig.
2(b), since the critical current $J'_c$ at the each microresistor
is slightly different.  This yields a small variation in
$\epsilon_i$ between the two defect sites (i.e., $\vert \epsilon_1
- \epsilon_2 \vert =\delta \epsilon$ and $\delta \epsilon \ll 1$).
The symmetric potential $V(q)$ shown in Fig. 2(a) has the
single-well structure for
$\ell<\ell_o=\ln[(\sqrt{3}+1)/(\sqrt{3}-1)]\approx 1.317$, but it
has the double-well structure for $\ell>\ell_o$. For
$\ell=\ell_o+a$, with $a \ll 1$, the fluxon potential $V(q)$ may
be expanded around $q=0$ to obtain
\begin{equation}
V(q)-V(0) \approx -{16 \epsilon \over 3\sqrt{3}}aq^2 + {32
\epsilon \over 27} q^4~. \label{double}
\end{equation}
The potential function of Eq. (\ref{double}) has the stable states
at $q=\pm q_o/2$ with $q_o=3[a/\sqrt{3}]^{1/2}$. The barrier
height $V_o$ between two stable states is $V_o=2\epsilon a^2$ and
the frequency $\omega_o$ of small oscillation around the stable
minimum is $\omega_o=\sqrt{(d^2 V(q)/d q^2)/M}=(8\epsilon
a/3\sqrt{3})^{1/2}$, indicating that $\omega_o \gg V_o$.  This
suggests that {\it the JVQ cannot be obtained by placing two
microresistors too closely (i.e., $\ell \approx \ell_o$) since the
potential barrier $V_o$ may not be strong enough to localize the
fluxon to either well}.  Hence a larger separation distance $\ell$
is needed to localize the bound states in either potential well.

\begin{figure}[t]
\includegraphics[width=6.5cm]{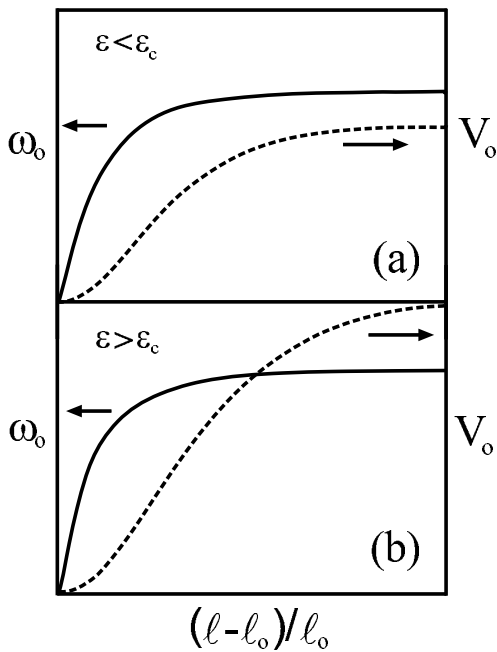}
\caption{ Dependance of the barrier potential $V_o$ (dashed line)
and the frequency of small oscillations $\omega_o$  (solid line)
for the symmetric double-well potential on the separation distance
$\ell$ between the two defect sites is schematically illustrated
for a) $\epsilon < \epsilon_c$ and b) $\epsilon > \epsilon_c$. The
critical pinning strength $\epsilon_c$ is 0.125.} \label{fig3}
\end{figure}

The barrier potential $V_o$ must be larger than $\omega_o$ in
order to obtain a localized ground state in either well without
mixing it with the excited state of the system. As $\ell$
increases from $\ell_o$, the value of both $V_o$ and $\omega_o$
increases, but this increase depends on $\epsilon$.  As $\ell
\rightarrow \infty$, $V_o$ approaches $2\epsilon$ while $\omega_o$
approaches $\sqrt{\epsilon/2}$.  This indicates that, when
$\epsilon$ is less than the critical value $\epsilon_c$ (i.e.,
$\epsilon < \epsilon_c$), $V_o$ remains smaller than $\omega_o$
for all $\ell$.  However, when $\epsilon > \epsilon_c$, $V_o$
becomes larger than $\omega_o$ as $\ell$ is increased, as shown
schematically in Fig. 3.  We estimate $\epsilon_c=0.125$, assuming
that $\omega_o(\epsilon_c )=V_o(\epsilon_c)$ at $\ell=\infty$.  We
will consider $\epsilon > \epsilon_c$ in the discussion below, so
that $V_o \geq \omega_o$.

For the fluxon localized at either left or right side of the
symmetric double-well shown in Fig. 2(a), the energy of the ground
state is degenerate.  We use eigenstates $\vert R \rangle$ and
$\vert L \rangle$ of the operator ${\hat \sigma}_z$ with
eigenvalues +1 and -1 to represent the right-localized and
left-localized state, respectively.  These two states are
exploited in the JVQ.  To do this, we need to ensure that the
tunneling rate between the two wells does not mix the ground state
with the excited states.

The fluxon in the ground state of the symmetric double-well
potential can tunnel from the left side to the right side (and
vice versa).  This MQT yields splitting of two degenerate fluxon
ground states. Within the semiclassical WKB
approximation,\cite{Lang} the tunneling rate $\Delta$ is given by
\begin{equation}
\Delta={\cal A}(0)~e^{-{\cal B}(0)}~, \label{tunnel1}
\end{equation}
where
\begin{eqnarray}
{\cal A}(0) &=& \left( {8\omega_o^3 q_o^2 \over \pi} \right)^{1
\over 2}~e^{2\int_0^{q_o \over 2}dq \left[ {\omega_o \over
\sqrt{{\bar V}(q)}} - {1 \over q_o-2q} \right]}~,
\label{tunnel2} \\
{\cal B}(0) &=& 4 \int_{-{q_o \over 2}}^{q_o \over 2} dq
~\sqrt{{\bar V}(q)}~, \label{tunnel3}
\end{eqnarray}
and ${\bar V}(q)=V(q)-V(\pm q_o/2)$ denotes the potential energy
measured from the bottom of the well.  The MQT in the real space
represents particle-like collective excitation, reflecting the
behavior of the fluxon as a quantum particle.  The computed
tunneling rate, using the WKB approximation, yields good agreement
with the quantum result when many states are bounded by the
double-well potential, but this agreement is poor when only the
ground state is bounded.  In Fig. 4, we compare the result of the
semiclassical WKB calculation using Eqs.
(\ref{tunnel1})-(\ref{tunnel3}) and the quantum mechanical
calculation to illustrate this difference.  The difference between
these two results is noteworthy: the WKB calculation overestimates
$\Delta$.  This difference is large when $\ell$ is small (i.e.,
small $V_o$) but decreases with increasing $\ell$ (i.e.,
increasing $V_o$).  We will use the quantum result for $\Delta$ in
the discussion below since the coherence time depends on $\Delta$.

\begin{figure}[t]
\includegraphics[width=6.9cm]{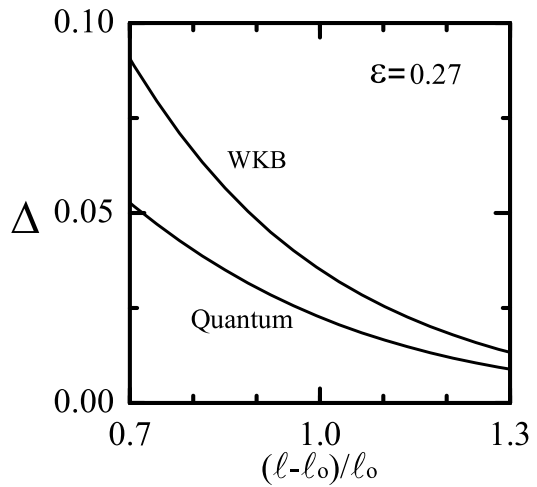}
\caption{ The splitting $\Delta$ of the ground state for
$\epsilon=0.27$ is plotted to compare the results obtained from
the WKB approximation and that obtained from the quantum
mechanical calculation. } \label{fig4}
\end{figure}

Numerical solution of the bound state energy for the potential of
Eq. (\ref{poten}) (with $f(t)=0$) indicates that the ground state
is localized at the either side of the double-well for only
limited value of $\ell$, as shown in Fig. 5.  The lower and upper
shaded areas in Fig. 5 represent the regions in the ($\epsilon$,
$\ell$) parameter space where the tunneling rate between the two
degenerate ground states and the excited states, respectively, are
large so that these states cannot be localized in either well. For
a fixed $\epsilon$, the number of localized states in either well
increases with $\ell$.  This indicates that the separation
distance $\ell$ and the pinning strength $\epsilon$ may be chosen
so that only the ground state is localized in either well. {\it
The parameters which yield localization of only the ground state
(i.e., between two shaded regions) may be ideal for obtaining the
JVQ.}

\begin{figure}[t]
\includegraphics[width=6.5cm]{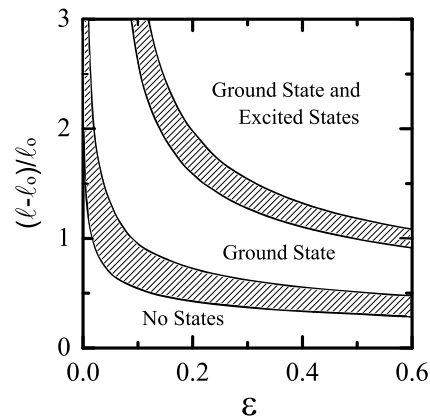}
\caption{ The diagram in the $(\ell,\epsilon )$ parameter space
illustrating that bound fluxon states localized in either side of
the double-well potential.  The shaded areas represent the region
with large tunneling rate. } \label{fig5}
\end{figure}

\section{spin-boson model}

In this section, we describe the interaction between the JVQ and
noisy environment.  We proceed by describing the fluxon dynamics
of Eq. (\ref{action}) in terms of the well-known spin-boson model.
This may be carried out by using the two-dimensional Hilbert space
spanned by the two degenerate ground states: the fluxon localized
at the left well (i.e., $\vert L \rangle$ ) and at the right well
(i.e., $\vert R \rangle$ ).  Following earlier studies, we
consider the parameter regime of $V_o \geq \omega_o \gg \Delta,
{\bar \epsilon}, T$ and include the effects of quasiparticle
dissipation and fluctuating weak bias current in the spin-boson
model.\cite{Legg,Weiss} The Hamiltonian for this model is written
as
\begin{equation}
H=H_S+H_{SB}+H_{B}~.
\label{Ham}
\end{equation}
The spin (S) Hamiltonian $H_S$,
\begin{equation}
H_S=-{1 \over 2} \Delta {\hat \sigma}_x -{1 \over 2}[{\bar
\epsilon} + {\bar f}(t)]{\hat \sigma}_z~,
\label{spin}
\end{equation}
describes the two-state qubit system, which is obtained from the
double-well potential of Eq. (\ref{poten}).  Here $\Delta$ is the
tunneling rate between the two wells.  The Pauli operators, ${\hat
\sigma}_z$ and ${\hat \sigma}_x$, in Eq. (\ref{spin}) represent
\begin{equation}
{\hat \sigma}_z = \vert R \rangle \langle R \vert - \vert L
\rangle \langle L \vert~,
\end{equation}
and
\begin{equation}
{\hat \sigma}_x = \vert R \rangle \langle L \vert + \vert L
\rangle \langle R \vert~,
\end{equation}
respectively.  The Hamiltonian $H_S$ also accounts for a
modification of the simple two-state system by a small asymmetry
in the potential due to slight variation in the pinning strength
of the microresistors (i.e., $\epsilon_1 \ne \epsilon_2$) and by
fluctuating bias current (i.e., $f(t)$).  The bias current density
${\bar f}(t)=q_o f(t)$, representing the driving force for the
fluxon, consists of two parts
\begin{equation}
f(t)=f_o + \delta f(t)~,
\label{biascurr}
\end{equation}
where $f_o$ and $\delta f (t)$ denote the homogeneous and randomly
fluctuating weak bias current components, respectively.  Here
$\delta f(t)$ accounts for the current noise in the JVQ.  The
Ohmic environment,\cite{CL} which accounts for the quasiparticle
dissipation, is described by the bath (B) Hamiltonian $H_B$,
\begin{equation}
H_B = {1 \over 2} \sum_{i=1}^N \left( {P_i^2 \over m_i} +
m_i\omega_i^2 Q_i^2 \right)~.
\label{bath}
\end{equation}
The interaction between the qubit system and the dissipative
environment is described by the spin-bath (SB) Hamiltonian
$H_{SB}$,
\begin{equation}
H_{SB} = - {\hat \sigma}_z{q_o \over 2} \sum_{i=1}^N c_i Q_i~.
\label{spinbath}
\end{equation}
It is noted that the spin-boson model of Eqs. (\ref{Ham}),
(\ref{spin}), (\ref{bath}), and (\ref{spinbath}) neglects the
contributions from the excited states that are bounded by the
potential well.  Hence, thermally activated leakage,\cite{leak}
which may also contribute to decoherence at finite $T$, is not
accounted in this work.  However, we may, safely, assume that this
contribution is negligible at ultra-low temperatures.
Consequently, the weakly fluctuating bias current (i.e., current
noise) at low frequency is the dominant source for dephasing at
these temperatures.

We now discuss the time dependent bias current $f(t)$ of Eq.
(\ref{biascurr}).  We set the externally applied homogeneous
component of the bias current to zero (i.e., $f_o=0$) since it
yields unwanted asymmetry in the double-well potential (see Fig.
2(b)) for the fluxon.  The weak bias current fluctuation ($\delta
f(t)$), representing random force in the LJJ due to nonequilibrium
states, yields small time-dependent asymmetry in the double-well
potential.  This asymmetry may be made small but cannot be turned
off completely as it arises from the noise-producing environment.
This current noise leads the basis states $\lbrace \vert L \rangle
, \vert R \rangle \rbrace$ to fluctuate weakly, as shown
schematically in Fig. 6.  However, this effect on the basis state
in the JVQ is expected to be smaller than that in other
superconducting qubits since the bias current is not used to
control the qubit state.  It is noted that the effect of noise in
the bias current for $f_o \not= 0$ has been investigated for the
phase qubit\cite{Mar} and for the charge qubit.\cite{Wei} In these
qubits, the bias current (i.e., $f_o$) is used to control the
qubit.  Noise in the bias current (i.e., $\delta f$) affects the
coherence time of the qubit since it leads to the fluctuation of
the qubit state.  The bias current noise leads to phase noise for
the phase qubit\cite{Mar} and radiation noise for the charge
quibit.\cite{Wei}

To model the effect of this random force more realistically in the
JVQ, we describe the bias current fluctuation as Gaussian colored
noise with non-zero characteristic correlation time $\tau_n$.  The
current noise $\delta f(t)$ has two main effects on the dynamics
of qubit density matrix: i) it leads to transition between two
energy eigenstates, and ii) it suppresses coherence between the
eigenstates by contributing to pure dephasing.  The Gaussian
colored noise can be used to account for the noise spectrum with
pronounced frequency dependence, such as Lorentzian noise and low
frequency asymmetrical magnetic field fluctuations\cite{MC} in the
tunnel junction.  The characteristics of the bias current are
described as
\begin{equation}
\langle \delta f(t) \rangle = 0 ~, \label{noise1}
\end{equation}
and
\begin{equation}
\langle \delta f(t)~\delta f(t') \rangle = n_o^2~ e^{-\vert t-t'
\vert /\tau_n}~. \label{noise2}
\end{equation}
Here $\langle \cdot \cdot \cdot \rangle$ denotes average over
different realizations of the fluctuating current, and $n_o$ is
the typical noise amplitude.  We note that, as $\tau_n \rightarrow
0$, the colored noise of Eq. (\ref{noise2}) becomes the white
noise, which is characterized by the correlation function $\langle
\delta f(t)~\delta f(t') \rangle = n_o^2 \delta(t-t')$.  The
correlation function of Eq. (\ref{noise2}) indicates that the
effects of the current noise for $t \ll \tau_n$ differ from those
for $t \gg \tau_n$.  For $t\ll \tau_n$, decay of coherence arises
from averaging over the distribution of current noise since the
fluctuations appear static.  For $t \gg \tau_n$, on the other
hand, decay of coherence is expected to be exponential since the
fluctuating bias current behaves as white noise.  The crossover
behavior occurs at $t \simeq \tau_n$.  The spectral density of the
bias current noise can be taken as
\begin{equation}
S_{noise}(\omega )={2n_o^2 \tau_n \over 1 + (\omega \tau_n)^2}~.
\label{noise3}
\end{equation}
The Lorentzian spectrum of Eq. (\ref{noise3}), characterizing
telegraph (diachotomous) noise, was observed in intrinsic
LJJ.\cite{IJJ}  We note that the noise spectrum in a small tunnel
junction is described by the Lorentzian function of Eq.
(\ref{noise3}), but the $1/\omega$-like noise spectrum in a larger
junction may be obtained as a result of several superimposed
Lorentzian features.\cite{noise}  In the discussion below, we make
few assumptions about the correlation function of Eq.
(\ref{noise2}): the fluctuation is weak and has small
characteristic amplitude (i.e., $n_o \ll \Delta$), but it has a
long correlation time (i.e., $\tau_n \gg 1/\Delta$). Also, we
assume that the temperature ($T$) of the bias current producing
environment is larger than the cutoff frequency of $1/\tau_n$
(i.e., $T \gg 1/\tau_n$).

\begin{figure}[t]
\includegraphics[width=6.5cm]{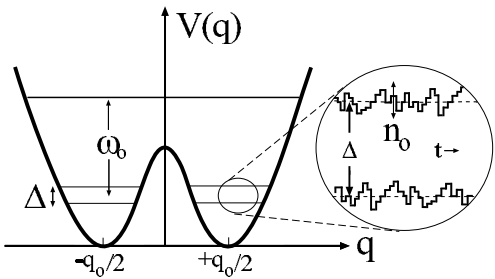}
\caption{A schematic diagram illustrating the effect of colored
noise on the bounded ground state of Josephson vortex in the
symmetric double-well potential.}
\label{fig6}
\end{figure}

\section{decoherence due to fluctuating weak bias current in Ohmic environment}

We now discuss the effect of dissipation and noisy environment on
the coherence time of the JVQ by using well-established formalism.
Here the coherence time represents the time scale for decay of
macroscopic quantum coherence (MQC) between the ground states in
the double-well potential. Here MQC is due to quantum tunneling of
the fluxon which leads to coherent oscillations. This MQC is
suppressed by the two decoherence sources since the interaction
between the qubit system and its environment can easily destroy
the phase coherence between two states.  In estimating $T_2$, we
follow the standard theoretical approach of using the
Bloch-Redfield theory and making lowest order Born approximation.
The effects of these two sources may be characterized as follows.
The Ohmic environment yields the finite relaxation time ($T_1$)
and dephasing time ($T_\phi^B$). However, the fluctuating weak
bias current modifies $T^B_\phi$ without significantly changing
$T_1$. We estimate the effects of the bias current noise,
restricting our consideration to $t \gg \tau_n$ since the bias
current fluctuation appears as $\delta$-function correlated (i.e.,
white noise), and the quantum coherence decays exponentially.
Hence $T_\phi$ may be expressed simply as
\begin{equation}
{1 \over T_\phi}={1 \over T^B_\phi} + {1 \over T^{noise}_\phi}
\end{equation}
where $T^{noise}_\phi$ is the dephasing time due to weak bias
current noise.  This indicates that the divergence in $T_\phi^B$
at ultra-low temperature may be cut off by $T_\phi^{noise}$. The
contributions from the higher-order Born correction and
non-Markovian effect, which are not included in the present work,
are small but yield power-law decay.\cite{power} These
contributions may also cut off the diverging $T^B_\phi$ due to
significantly reduced interaction between the qubit system and
environment at ultra-low temperatures.

The decay of coherent oscillations is estimated by considering the
generalized master equation for the system's density matrix
$\rho_S (t)$,
\begin{equation}
{d \rho_S(t) \over dt}= -i [H_S, \rho_S(t)] -i\int_0^t dt'
\Sigma_S(t-t') \rho_S(t')~,
\label{master}
\end{equation}
and by assuming that the time dependence of the decoherence source
is weak.  Here $\rho_S(t)={\rm Tr}_B \rho(t)$, $\rho=\rho_S
\bigotimes \rho_B$, the kernel $\Sigma_S (t)$ is the self-energy
operator
\begin{equation}
\Sigma_S(t)= -i {\rm Tr}_B {\cal H}_{SB} e^{-{\cal QH}t} {\cal
H}_{SB}~\rho_B~,
\end{equation}
${\cal H}_{SB}$ is the Liouvillian superoperator defined by ${\cal
H}_{SB}~\rho = [H_{SB},\rho]$, and  ${\cal Q}=1-\rho_B {\rm Tr}_B$
is the projection superoperator.   Here $\rho_B$ is the bath
density matrix. Since the studies\cite{Hart} indicate that both
Bloch-Redfield theory and path integral theory yield equivalent
results, we employ the former approach for convenience.

The generalized master equation within the Born approximation is
obtained by using the fact that the coupling between the qubit
system and environment, as described by $J_\beta(\omega)/\omega
\propto \beta$, is small at low temperatures.  We follow the
Redfield theory\cite{redfield} and make a systematic perturbation
expansion of the kernel $\Sigma_S$ in powers of the system-bath
coupling ($\beta$).  We retain only the lowest order terms in this
expansion. Replacing $e^{-i{\cal QH}t} \rightarrow e^{-i{(\cal
H}_S+{\cal H}_B)t}$ and keeping the expansion of the kernel
$\Sigma_S$ up to the second order in $H_{SB}$, we obtain
\begin{equation}
\Sigma_S^{(2)}(t)= -i {\rm Tr}_B~ {\cal H}_{SB} e^{-i({\cal H}_S +
{\cal H}_B)t} {\cal H}_{SB}~\rho_B~.
\end{equation}
Further simplification of Eq. (\ref{master}) may be made by
assuming Markov system dynamics
\begin{equation}
\rho_S(t-\tau_o) \sim e^{i{\cal H}_S \tau_o}~\rho_S(t)~,
\end{equation}
in which the temporal correlation time $\tau_o$ in the dissipative
environment is very short due to very short-lived system-bath
interactions, and the bath correlation function decays to zero at
a very short-time.

{\it Bloch-Redfield equation:} In examining the decoherence
effects, the Hilbert space spanned by the ground states of the two
wells (Fig. 2(a)) is not convenient since the spin Hamiltonian of
Eq. (\ref{spin}) is not diagonal in the basis $\{ \vert L \rangle
, \vert R \rangle \}$.  We represent the two-state system in new
basis $\{ \vert 0 \rangle , \vert 1 \rangle \}$ given by
\begin{eqnarray}
\vert 0 \rangle &=& -\vert L \rangle \sin\theta + \vert R \rangle
\cos\theta ~,
\\
\vert 1 \rangle &=& ~~\vert L \rangle \cos\theta + \vert R \rangle
\sin\theta~,
\end{eqnarray}
where $\sin\theta =\sqrt{(\Omega-{\bar
\epsilon})/\Omega}/\sqrt{2}$,  $\cos\theta =\sqrt{(\Omega+{\bar
\epsilon})/\Omega}/\sqrt{2}$, and $\Omega=\sqrt{{\bar \epsilon}^2+
\Delta^2}$.  In this new basis $\lbrace \vert 0 \rangle, \vert 1
\rangle \rbrace$, we estimate $T_2$ by making the Born-Markov
approximation and by obtaining the Bloch-Redfield equations.
Taking matrix elements in the eigenbasis $\vert n \rangle$ of
$H_S$ (i.e., $\vert 0 \rangle$ and $\vert 1 \rangle$), we may
write the Redfield equations as
\begin{equation}
{d \rho_{nm}^S(t) \over dt}=-iE_{nm}~\rho_{nm}^S(t) - \sum_{kl}
{\cal R}_{nmkl}~\rho_{kl}^S(t)~, \label{Redfield}
\end{equation}
where $\rho_{nm}^S=\langle n \vert \rho_S \vert m \rangle$, ${\cal
R}_{nmkl}$ is the Redfield tensor, $E_{nm}=E_n -E_m$, and $E_n$ is
the eigenstate energy of $H_S$ (i.e., $H_S \vert n \rangle = E_n
\vert n \rangle$).   In the absence of the fluctuating bias
current ($\delta f$), the eigenstate energies in the diagonal
basis, representing the ground state splitting, are expressed as
\begin{eqnarray}
E_0 &=& -{1 \over 2} \sqrt{{\bar \epsilon}^2 + \Delta^2} ~,\\
E_1 &=& +{1 \over 2} \sqrt{{\bar \epsilon}^2 + \Delta^2}~.
\end{eqnarray}
In the presence of the low frequency bias current fluctuations,
these energies fluctuate slowly as shown schematically in Fig. 6.
We assume that the eigenstate energies are almost constant in the
time scale relevant for the evolution of the density matrix.

The Redfield tensor ${\cal R}_{nmkl}$ is defined by
\begin{eqnarray}
{\cal R}_{nmkl}=~~~~~~~~~~~~~~~~~~~~~~~~~~~~~~~~~~~~~~~~~~~~~~~~~~
\nonumber \\
\int_0^\infty dt {\rm Tr}_B \langle n \vert [ H_{SB}(t),[
H_{SB}(0), \vert k(t)\rangle \langle l(t) \vert \rho_B ] ] \vert m
\rangle \label{redtensor}
\end{eqnarray}
where the spin-bath Hamiltonian $H_{SB}$ and qubit system
eigenstate $\vert k \rangle$ in the interaction picture are
written as
\begin{eqnarray}
H_{SB}(t)&=&e^{i(H_S+H_B)t} H_{SB} e^{-i(H_S+H_B)t}~, \\
\vert k(t) \rangle &=& e^{itH_S}\vert k \rangle = e^{itE_k} \vert
k \rangle~,
\end{eqnarray}
respectively.  The Redfield tensor may be expressed as
\begin{equation}
{\cal R}_{nmkl}=\delta_{lm}\sum_r\Gamma_{nrrk}^{(+)} + \delta_{nk}
\sum_r\Gamma_{lrrm}^{(-)} - \Gamma_{lmnk}^{(+)} -
\Gamma_{lmnk}^{(-)}
\end{equation}
by evaluating the commutators in Eq. (\ref{redtensor}).  Here
\begin{eqnarray}
\Gamma_{lmnk}^{(+)}&=&\int_0^\infty dt~e^{-it\omega_{nk}}{\rm
Tr}_B {\bar H}_{lm}^{SB}(t) {\bar H}_{nk}^{SB}(0)~\rho_B, \\
\Gamma_{lmnk}^{(-)}&=&\int_0^\infty dt~e^{-it\omega_{lm}}{\rm
Tr}_B {\bar H}_{lm}^{SB}(0) {\bar H}_{nk}^{SB}(t)~\rho_B,
\end{eqnarray}
and ${\bar H}_{nm}^{SB}(t)=\langle n \vert e^{itH_B}H_{SB}
e^{-itH_B} \vert m \rangle$.  The relation
\begin{equation}
(\Gamma_{lmnk}^{(+)})^*=\Gamma_{knml}^{(-)}
\end{equation}
may be used to write the Redfield tensor in terms of only the
complex $\Gamma_{lmnk}^{(+)}$ tensor
\begin{eqnarray}
\Gamma_{lmnk}^{(+)}&&={2q_o^2} \langle l \vert {\hat \sigma}_z
\vert m \rangle \langle n \vert {\hat \sigma}_z \vert k \rangle
\int_0^\infty {dt \over 2\pi}~ e^{-iE_{nk}t} \times
\nonumber \\
&&\int_0^\infty d\omega J_\beta (\omega ) \left[ \coth{\omega
\over 2T} \cos \omega t - i \sin \omega t \right] \label{Gam}
\end{eqnarray}
with the spectral function $J_\beta (\omega)$ of Eq. (\ref{Spec})
representing Ohmic environment.

The dynamics of the two-state system may be described by using a
2-by-2 density matrix which is written in the Bloch vector form
(i.e., three real variables).  The Bloch vector $\vec p$ is
written as
\begin{equation}
{\vec p}={\rm Tr}({\vec \sigma}~\rho_S) = \left( \matrix{
\rho_{01}^S+\rho_{10}^S \cr i(\rho_{01}^S-\rho_{10}^S) \cr
\rho_{00}^S-\rho_{11}^S \cr } \right)=\left( \matrix{ \rho_+^S \cr
\rho_-^S \cr \rho_z^S \cr} \right) \label{bloch}
\end{equation}
where ${\vec \sigma}=(\sigma_x, \sigma_y, \sigma_z)$ represents
the vector composed of the three Pauli matrices.  The Bloch vector
of Eq. (\ref{bloch}) may be combined with the Redfield equation of
Eq. (\ref{Redfield}) to obtain the Bloch-Redfield equation
\begin{equation}
{d{\vec p} \over dt}={\vec e} \times {\vec p} - R {\vec p} + {\vec
p}_o ~, \label{BR}
\end{equation}
where ${\vec e}=(0,0,E_{01})^T$, the relaxation matrix $R$ is
given by
\begin{equation}
R= \left( \matrix{ {\cal R}'_{0101} + {\cal R}'_{0110} & {\cal
R}^{''}_{0101} - {\cal R}^{''}_{0110} & {\cal R}'_{0100} - {\cal
R}'_{0111} \cr -{\cal R}^{''}_{0101} - {\cal R}^{''}_{0110} &
{\cal R}'_{0101} - {\cal R}'_{0110} & {\cal R}^{''}_{0111} - {\cal
R}^{''}_{0100} \cr 2{\cal R}'_{0001} & 2{\cal R}^{''}_{0001} &
{\cal R}'_{0000} + {\cal R}'_{1111} \cr } \right)
\end{equation}
and
\begin{equation}
{\vec p}_o = \left( \matrix{ -({{\cal R}'_{0111}} + {{\cal
R}'_{0100}} ) \cr {\cal R}^{''}_{0100} + {\cal R}^{''}_{0111} \cr
-( {{\cal R}'_{0000}} - {{\cal R}'_{1111}} ) \cr } \right) ~.
\end{equation}
Here ${\cal R}'_{nmkl}$ and ${\cal R}^{''}_{nmkl}$ are the real
and the imaginary part of the Redfield tensor, respectively.

The Bloch-Redfield equation of Eq. (\ref{BR}) may be simplified
within the secular approximation.  Within this random phase type
approximation which corresponds to retaining only the terms ${\cal
R}_{nmkl}$ with the indices $n-m=k-l$, the Redfield tensor
simplifies to $R \simeq R_{sec}$ and
\begin{equation}
R_{sec} = \left( \matrix{ {\cal R}'_{0101} & {\cal R}^{''}_{0101}
& 0 \cr -{\cal R}^{''}_{0101} & {\cal R}'_{0101} & 0 \cr 0 & 0 &
{\cal R}'_{0000}+{\cal R}'_{1111} \cr } \right)~.
\end{equation}
This approximation is valid for the spin-boson model of Eq.
(\ref{Ham}).  We note that $E_{01}=\Omega$ while ${\cal R}_{nmkl}
\leq {\cal O}(\beta)$, when $n-m \ne k-l$.  Since $\Omega \gg
\beta$ at ultra-low temperatures, $E_{01} \gg {\cal R}_{nmkl}$.

{\it Ohmic environment:} We solve the Bloch-Redfield equation of
Eq (\ref{BR}) within the secular approximation to obtain the
relaxation and dephasing time due to the environment.  The decay
of the diagonal element of the qubit's reduced density matrix is
written as
\begin{equation}
\rho^S_z(t)=\rho^S_z(0)~e^{-t/T_1}~.
\end{equation}
This yields the relaxation time ($T_1$) which is given by
\begin{equation}
{1 \over T_1}={\cal R}'_{0000}+{\cal R}'_{1111}=2 {\rm
Re}(\Gamma^{(+)}_{0110} + \Gamma^{(+)}_{1001})~.
\end{equation}
Evaluating the tensors $\Gamma^{(+)}_{0110}$ and
$\Gamma^{(+)}_{1001}$ by using Eq. (\ref{Gam}), we obtain the
relaxation time $T_1$ as
\begin{equation}
{1 \over T_1}={4\Delta^2 \over \Omega}~q_o^2\beta~\coth {\Omega
\over 2T}~.
\label{relax}
\end{equation}
This is consistent with the result from the bounce
solution.\cite{bounce} The decay of the off-diagonal element of
the reduced density matrix (either $\rho^S_{01}$ or $\rho^S_{10}$)
may be expressed as
\begin{equation}
\rho^S_{01} (t)=\rho^S_{01} (0) {\cal I}(t)~e^{-t/T^B_2}~e^{+
i[\Omega +{\cal R}^{''}_{0101}]t}~. \label{offdiag}
\end{equation}
This reduced density matrix includes the effects from both the
Ohmic environment (i.e., $T_2^B$) and the fluctuating weak bias
current (i.e., ${\cal I}(t)$). The coherence time $T^B_2$ due to
Ohmic environment is obtained as
\begin{eqnarray}
{1 \over T^B_2} &=& {\rm Re}(\sum_r\Gamma^{(+)}_{0rr0} + \sum_r
\Gamma^{(-)}_{1rr1} - \Gamma^{(+)}_{1100} - \Gamma^{(-)}_{1100})~
\nonumber \\
&=& {1 \over 2T_1} + {1 \over T^B_\phi}~.
\label{dephaseen}
\end{eqnarray}
From Eq. (\ref{dephaseen}), it is straightforward to obtain the
dephasing time ($T^B_\phi$) due to the Ohmic environment as
\begin{equation}
{1 \over T_\phi^B}={16 {\bar \epsilon}^2 \over \Omega^2}~q^2_o
\beta~T~.
\label{deprev}
\end{equation}
This indicates that $T_\phi^B$ diverges as either ${\bar
\epsilon}$ or $\beta$ vanishes.  We note that ${\bar \epsilon}$ is
a temperature independent parameter, but $\beta$ becomes
exponentially small at ultra-low temperatures, yielding strong
divergence in $T_\phi^B$.

{\it Fluctuating weak bias current:}  The dephasing time
$T_\phi^{noise}$ due to the weak bias current noise may cut off
the divergent $T_\phi^B$ at ultra-low temperatures. We estimate
$T^{noise}_\phi$ from ${\cal I}(t)$ of Eq. (\ref{offdiag}).  The
suppression factor ${\cal I}(t)$ in the off-diagonal element of
reduced density matrix $\rho^S_{01}$ represents the
decay\cite{RSA} of coherence due to the bias current noise. This
suppression factor
\begin{equation}
{\cal I}(t) = \exp \bigg\{ \pm i \int_0^t dt'\left[ {{\bar
\epsilon} {\bar f}(t') \over \Omega} + {\Delta^2 {\bar f}^2(t')
\over 2 \Omega^3} \right] \bigg\}
\end{equation}
accounts for the accumulation of the noise induced phase between
two instantaneous energy eigenstates $\vert 0 \rangle$ and $\vert
1 \rangle$, due to long correlation time $\tau_n$. We estimate
$T^{noise}_\phi$ by averaging ${\cal I}(t)$ over the realization
of the fluctuating bias current and obtain
\begin{equation}
I(t) = \bigg\langle \exp \bigg\{ \pm i \int_0^t dt'\left[ {{\bar
\epsilon} {\bar f}(t') \over \Omega} + {\Delta^2 {\bar f}^2(t')
\over 2 \Omega^3} \right] \bigg\} \bigg\rangle~. \label{funcave}
\end{equation}
Here $\langle \cdot\cdot\cdot \rangle$ denotes the average over
noise realization.   For simplicity, we assume that the
fluctuating bias current is described as a Gaussian noise with the
correlation function of Eq. (\ref{noise2}) and the spectral
density of Eq. (\ref{noise3}).  We represent the average $\langle
\cdot\cdot\cdot \rangle$ by writing it as a functional integration
over the noise.  The transition probability $\cal P$ between
different noise realizations may be described by the Fokker-Planck
equation for the Ornstein-Uhlenbeck process\cite{FPOU}
\begin{equation}
{\partial {\cal P} \over \partial t} ={1 \over \tau_n} {\partial
\over \partial {\bar f}}({\bar f}{\cal P}) + {n_o^2 \over \tau_n}
{\partial^2 \over
\partial {\bar f}^2} {\cal P}~.
\end{equation}
For this process, the transition probability ${\cal P}({\bar
f},t;{\bar f}',t')$ for the noise from the value ${\bar f}$ at
time $t$ to the value ${\bar f}'$ after a time $\delta t = t-t'$
is given by
\begin{eqnarray}
{\cal P}({\bar f},t;{\bar f}',t')&&= \left[ {2\pi n_o^2} \left(
1-e^{-2\delta t/\tau_n} \right)\right]^{-1/2} \times \nonumber \\
&& \exp\bigg\{ -{1 \over 2n_o^2} {[{\bar f}-{\bar f}' e^{-\delta
t/\tau_n}]^2 \over 1 - e^{-2\delta t/\tau_n}} \bigg\} .
\label{prodis}
\end{eqnarray}
We use this transition probability to express the probability of
specific noise realization as
\begin{equation}
{\cal P}_o({\bar f}_0){\cal P}({\bar f}_0,0;{\bar f}_1,t_1){\cal
P}({\bar f}_1,t_1;{\bar f}_2,t_2)\cdot\cdot\cdot {\cal P}({\bar
f}_{n-1},t_{n-1};{\bar f}_t,t)
\end{equation}
where ${\bar f}_i={\bar f}(t_i)$  and ${\cal P}_o({\bar f})=(2\pi
n_o^2)^{-1/2} \exp (-{\bar f}^2/2n_o^2)$ is the stationary
Gaussian probability distribution of ${\bar f}$.  We note that
$\delta t_i=(t_i-t_{i-1})/n$.  In the limit of $\delta t_i
\rightarrow 0$ (i.e., $n \rightarrow \infty$), the average over
the noise realization may be expressed as
\begin{eqnarray}
\bigg\langle \cdot\cdot\cdot \bigg\rangle &=& \left({1\over 2\pi
n_o^2}\right)^{1 \over 2} e^{t \over 2\tau_n} \int d{\bar f}_0~
d{\bar f}_t~D[{\bar f}(t')]~\cdot\cdot\cdot~\times \nonumber \\
&&~~~e^{ -{({\bar f}^2_0 +{\bar f}^2_t)\over 4n_o^2} - \int_0^t
{dt' \over 4n^2_o\tau_n} \left[\tau_n^2 \left( {d{\bar f} \over
dt} \right)^2 + {\bar f}^2 \right] }~, \label{noise}
\end{eqnarray}
where $D[{\bar f}]=\prod_{i=1}^{n-1}\{d{\bar f}_i /[4\pi n^2_o
\sinh(\delta t_i/\tau_n)]^{1/2}\}$ denotes the measure.  The
functional integral of Eqs. (\ref{funcave}) and (\ref{noise}) is
similar to that for a driven harmonic oscillator.\cite{FH} Using
this similarity, we may carry out the average over the realization
of fluctuating bias current straightforwardly and obtain
\begin{equation}
I({\bar t})=I_o({\bar t}) \exp \bigg \{ -b_o \left[ {{\bar \omega}
{\bar t}} - {2 \over \coth \left( {{\bar\omega} {\bar t} \over
2}\right) + {\bar\omega}} \right] \bigg \}
\end{equation}
where ${\bar t}=t/\tau_n$,
\begin{equation}
I_o({\bar t})=e^{{\bar t}\over 2} \left[ \cosh \left( {\bar\omega}
{\bar t} \right) + {1+{\bar\omega}^2 \over 2{\bar\omega}} \sinh
\left( {\bar\omega} {\bar t} \right) \right]^{-{1 \over 2}} ,
\end{equation}
${\bar\omega}=\sqrt{1+2i{\bar n}_o^2 {\bar\tau_n}
(\Delta/\Omega)^3}$, and $b_o={\bar \epsilon}^2 {\bar n}_o^2 {\bar
\tau_n}^2/(\Omega^2/{\bar \omega}^3)$.  Here the two dimensionless
parameters, ${\bar n}_o=n_o/\Delta$ and ${\bar
\tau_n}=\tau_n\Delta$, characterize the amplitude and correlation
time for the fluctuating bias current, respectively.  For ${\bar
t} \ll 1$, the fluctuation appears static.  Hence the average
$\langle \cdot\cdot\cdot \rangle$, which is over the static
distribution of noises, yields
\begin{equation}
I({\bar t})= \left( {1 + {\bar t} \over 1+ {\bar t}+i2{\bar
n}_o^2{\bar \tau}_n {\bar t}} \right)^{1 \over 2} \exp \bigg \{-
 {{\bar \epsilon}^2 {\bar n}_o^2 {\bar \tau}_n^2
{\bar t}^2 \over 2 \Omega^2} \bigg \}~.
\end{equation}
For ${\bar t} \gg 1$, on the other hand, the fluctuation appears
to be $\delta$-function correlated (i.e., white noise), and hence,
we obtain that $I(t) \propto \exp (-t/T_\phi^{noise})$. This
exponential decay indicates that the dephasing time
$T_\phi^{noise}$ may be expressed as
\begin{eqnarray}
{1 \over T_\phi^{noise}}&=&{\Delta \over 2\sqrt{2} {\bar \tau}_n}
\Bigg [ \left( \sqrt{1 + 4{\bar n}_o^4 {\bar\tau}_n^2 {\Delta^6
\over \Omega^6}} +1 \right)^{1 \over 2} - \sqrt{2}
\nonumber \\
&& + {{\bar \epsilon}^2 \over \Omega^2}{ 2\sqrt{2} {\bar
n}_o^2{\bar\tau}_n^2 \over \sqrt{1+ 4{\bar n}_o^4
{\bar\tau}_n^2{\Delta^6 \over \Omega^6} }} \Bigg ]~.
\label{depnoise}
\end{eqnarray}
We note that $T_\phi^{noise}$ is independent of $T$. Consequently,
this will eventually cut off the divergent $T_\phi^B$ due to small
coupling between the qubit system and environment at ultra-low
$T$.  Also, the effects due to the bias current is reduced, as
expected, when asymmetry in the double-well potential vanishes
(i.e., ${\bar \epsilon}=0$).

\section{Discussion}

We now estimate numerically the coherence time for the JVQ and
show that ultra-long coherence time can be obtained by using the
experimental value for the parameters.  Since a long Nb-AlO$_x$-Nb
junction may be used to fabricate the qubit, we use the following
experimental values in estimating $T_2$:\cite{Sakai,Kl}
$\lambda_L\sim$90 nm, $\lambda_J \sim$25 $\mu$m, $J_c \sim
$2$\times 10^6$ A/m$^2$ and $\omega_p\sim$90 GHz.  For
definiteness, we chose a narrow width (i.e. $L_y \sim$0.2 $\mu$m)
for the junction so that the quantum effect is enhanced.  Also, we
set that ${\bar \epsilon}/ \Delta=0.01$ since the variation in the
pinning strength (i.e., $\bar \epsilon$) for the two defect sites
can be made small.  Here we use the tunneling rate $\Delta$ which
is obtained from quantum calculation of the ground state
splitting, as shown in Fig. 4.

As the coherence time $T_2$ depends strongly on $\Delta$, we
discuss, first, the dependence of $\Delta$ on the separation
distance $\ell$ between two defect sites and the pinning strength
$\epsilon$. In Fig. 7, the tunneling rate for the ground state of
the symmetric double-well potential (i.e., ${\bar \epsilon}=0$) is
plotted as a function of $\ell$ for $\epsilon=$0.21 (dashed line),
0.27 (solid line), and 0.33 (dot-dashed line). The curves show
that MQT in LJJ depends strongly on both $\ell$ and $\epsilon$, as
indicated by earlier studies.\cite{KM}  The decrease in $\Delta$
with increasing $\ell$ and/or $\epsilon$ reflects that the
tunneling rate decreases with increasing barrier potential $V_o$.
We use this numerical result, below, in estimating $T_\phi$.

\begin{figure}[t]
\includegraphics[width=6.9cm]{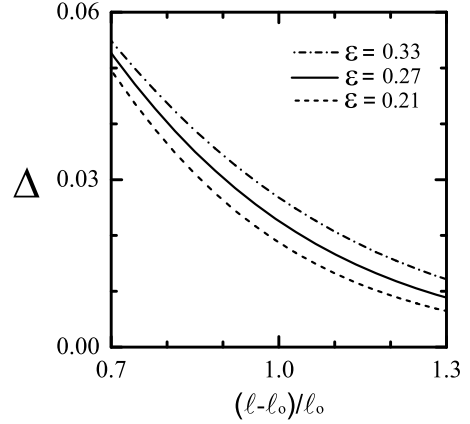}
\caption{ The dependance of the tunneling rate on the defect
separation distance $\ell$ and on the pinning strength $\epsilon$
is illustrated.}
\label{fig7}
\end{figure}

{\it Ohmic environment:} The relaxation time $T_1$ and the
dephasing time $T_\phi^B$ due to the interaction between the qubit
system and environment are estimated by using Eqs. (\ref{relax})
and (\ref{deprev}), respectively.  These two characteristic times
depend strongly on the quasiparticle dissipation effect (i.e.,
$\beta$). For definiteness, we chose $\epsilon=0.27$ and $T=25$
mK.  This ultra-low $T$ is chosen because the experiments
show\cite{Wa} that the localized fluxon behaves as a quantum
particle.  For $(\ell-\ell_o)/\ell_o=1.0$, the estimated values
are $T_1 \approx (0.005/\beta)$ ns and $T_\phi^B \approx
(28/\beta)$ ns, indicating that both $T_1$ and $T_\phi^B$ become
divergently long since $\beta$ is strongly reduced at this
temperature.  The estimated value\cite{Sakai} for $\beta$ is
roughly 0.03 at $T\sim 4$ K, and it is found to decrease
exponentially with $T$,\cite{PW} below the superconducting
transition temperature.  Phenomenologically,\cite{PW} the
dissipation effect represents the losses due to the tunnel
barriers.  These losses are related to the quasiparticle
resistance $R_{qp}(T)$ as
\begin{equation}
\beta= {1 \over \omega_p C R_{qp}(T)}
\label{beta}
\end{equation}
where $C$ is the capacitance associated with the tunnel barrier.
The $T$ dependence of the quasiparticle resistance below the
superconducting gap energy $\Delta_{sc}(T)$ is given by
\begin{equation}
R_{qp}(T)=R_T~e^{\Delta_{sc}(T)/T}
\end{equation}
where $R_T$ is the normal state tunneling resistance.  The
exponential $T$ dependence for $\beta$, as indicated in Eq.
(\ref{beta}) has been observed to low temperatures\cite{PW} (i.e.,
$T \ll T_c$).  We note that the dissipation coefficient $\beta_s$,
which represents the contribution from the quasiparticle current
along the junction layer, also decreases exponentially with
$T$,\cite{qpc} but this contribution is not included in this work.
This exponential $T$ dependence for $\beta$ suggests that both
$T_1$ and $T_\phi^B$ become divergently long at $T=25$ mK since
$T_1 \rightarrow 0$ and $T_\phi^B \rightarrow 0$ as $T \rightarrow
0$.  Both $T_1$ and $T_\phi^B$ are cut off by the decoherence due
to weak bias current fluctuations.  This suggests that the
measured coherence time $T_2$ at $T=25$ mK may be estimated as
$T_2 \approx T_\phi^{noise}$ since $1/T_2 \simeq
1/T_\phi^{noise}$.

{\it Fluctuating weak bias current:} The dephasing time $T_\phi$
is limited by the contribution due to the fluctuating bias current
(i.e., current noise).  As indicated in Eq. (\ref{depnoise}), the
dephasing time due to the bias current fluctuation depends on the
spectral density $S_{noise}(0)=2\Delta {\bar n}_o^2 {\bar \tau_n}$
of Eq. (\ref{noise3}).  The value of the spectral density
$S_{noise}(0)$ may be estimated by using the line-width, $\delta
\omega_{FFO}$, data for the Nb-AlO$_x$-Nb flux flow oscillator
(FFO).  In estimating $S_{noise}(0)$ we may use the relation
between the magnetic field and bias current fluctuations.  Recent
measurements\cite{Kos} of the line-width in FFO indicate that
fluctuating bias current $\delta f(t)$ in the control line
generates magnetic field fluctuation $\delta B(t)$ in the LJJ. In
reverse, magnetic field fluctuations due to both external and
internal sources produce bias current fluctuations.\cite{Pan} This
suggests that, when magnetic field in the LJJ fluctuates,
dephasing due to the bias current fluctuation may arise.  Since
the fluctuations are wide-band noises and are small, the relation
between the bias current and magnetic field fluctuations may be
expressed as
\begin{equation}
\delta B(t) = K~\delta f(t)~,
\end{equation}
where $K$ is the parameter of the order unity, describing
conversion between the bias current and magnetic field
fluctuations.  We note that $K$ depends on the geometry of the
LJJ.  The relation between the line-width $\delta\omega_{FFO}$ and
the spectral density $S_{noise}(0)$ for the current noise is
given\cite{Pan} by
\begin{equation}
\delta \omega_{FFO}={2\pi \over \Phi_o^2} (R_B + K R_H)^2
S_{noise}(0)~,
\end{equation}
where $\Phi_o$ is the flux quantum in the superconducting state,
$R_B$ is the differential resistance associated with the bias
current, and $R_H$ is the differential resistance associated with
the magnetic field.   The current noise spectral density measured
from the FFO is $S_{noise}^{FFO}(0) \simeq 2.8\times 10^{-22}$
$C^2/s$.  Accounting for the geometry of the LJJ used in the
experiment, the spectral density $S_{noise}^{FFO}(0)$ is given by
$S_{noise}^{FFO}(0)=(2.5\times 10^{-19}~C^2/s) S_{noise}(0)$,
yielding $S_{noise}(0) \simeq 0.0011$.  Here we used the following
parameters that are obtained from the experimental data for the
FFO:\cite{Kos} $R_B \simeq 0.03~\Omega$, $R_H \simeq
0.005~\Omega$, and $K\simeq 1$. Also we chose $\delta \omega_{FFO}
\simeq 500$ kHz, for definiteness, since the data indicate that
the FFO line-width at the plateau of Fiske steps does not decrease
below few hundred kHz at low values of $R_B$.

In Fig. 8, we plot the dephasing time $T_\phi^{noise}$ versus the
defect separation distance $\ell$ for $\epsilon=0.21$ (dashed
line), 0.27 (solid line), and 0.33 (dot-dashed line). Here
$T_\phi^{noise}$ is computed from Eq. (\ref{depnoise}) by using
the experimental value of $S_{noise}(0)\simeq 0.0011$ and by
assuming, for definiteness, that $\tau_n=1000$ (Fig. 8a) and
$\tau_n=500$ (Fig. 8b), which correspond to 10 ns and 5 ns,
respectively. The curves indicate that $T_\phi^{noise}$ increases
with $\tau_n$. Also, $T_\phi^{noise}$ decreases and increases with
$\ell$ and $\epsilon$, respectively, but it depends strongly on
$\ell$ and weakly on $\epsilon$.  The decrease in $T_\phi^{noise}$
with $\ell$ is due to the decrease in the tunneling rate, as shown
in Fig. 7.  We note that for $(\ell-\ell_o)/\ell_o=1.3$ the
condition of ${\bar \tau}_n \gg 1$ is becoming difficult to
satisfy due to small tunneling rate. For $\tau_n=500$ and
$(\ell-\ell_o)/\ell_o=1.3$, the computed value for ${\bar \tau}_n$
is 5.8, 4.2, and 3.1 for $\epsilon=0.33$, 0.27, and 0.21,
respectively.  For a smaller defect separation distance, say
$(\ell-\ell_o)/\ell_o=1.0$, the condition of ${\bar \tau}_n \gg 1$
is more easily satisfied and the computed value for
$T_\phi^{noise}$ is in the microsecond range for both
$\tau_n=1000$ and 500.  For $\epsilon=0.27$, $T_\phi^{noise}$ is
roughly 55 $\mu s$ and 30 $\mu s$ for $\tau_n=1000$ and 500,
respectively.   We compare this result with $T_\phi$ for other
superconducting qubits.  The measured values of $T_\phi$ are 20 ns
for the flux qubit\cite{flux} at 25 mK, 10 ns for the phase
qubit\cite{phase} at 25 mK, and 500 ns for the
quantronium\cite{chfl} at 15 mK.   These values indicate that
$T_\phi^{noise}$ for the JVQ is orders of magnitude larger than
the observed dephasing time in other qubits.  This difference
represents the fact that the fluxon's coupling to noisy
environment is substantially weaker than other superconducting
qubits.  This indicates that longer coherence time may be obtained
as fluctuating magnetic field in the LJJ is further reduced.
Moreover, phenomenological comparison of these qubits
indicates\cite{Har} that the ultra-long dephasing time may also be
attributed\cite{Har} to the fact that the JVQ has much larger
junction area than other qubits.

\begin{figure}[t]
\includegraphics[width=7.5cm]{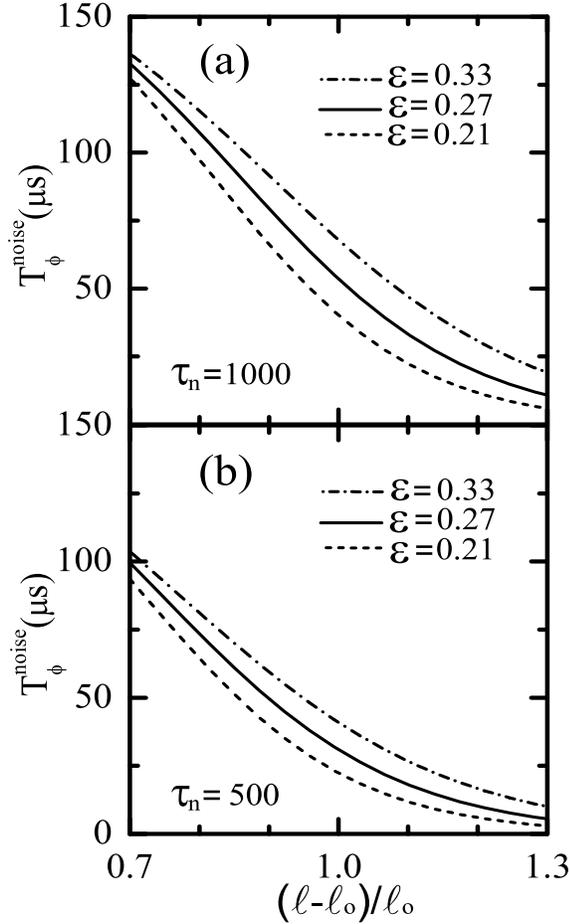}
\caption{ The dephasing time $T_\phi^{noise}$ versus the defect
separation distance $\ell$ is plotted for $\epsilon$=0.21 (dashed
line), 0.27 (solid line), and 0.33 (dot-dashed line) to illustrate
the dependence of $T_\phi^{noise}$ on both $\ell$ and $\epsilon$.
The computed $T_\phi^{noise}$ for a) $\tau_n=1000$ and b) 500
illustrate the dependence on the correlation time.} \label{fig8}
\end{figure}

\section{Summary and conclusion}

In summary, we investigated the coherence time for the JVQ which
may be fabricated by using a long Nb-AlO$_x$-Nb juntion.   Since
the critical current fluctuation does not contribute to dephasing
of the JVQ system, we estimate the coherence time by accounting
for two sources of decoherence: i) quasiparticle dissipation and
ii) current noise in the junction.  We note that, within the
lowest order approximation, the low frequency noise due to
critical current fluctuation does not couple to the JVQ, and
consequently it does not contribute to dechoerence.  However, the
low frequency noise due to bias current fluctuation is an
important decoherence source.  We showed that $T_1$ and $T_\phi^B$
due to the quasiparticle dissipation (i.e., Ohmic environment)
diverge at ultra-low temperatures (i.e., $\sim$ 25 mK) since the
dissipation effect (i.e., $\beta$) becomes exponentially small for
$T$ below the superconducting transition temperature.  In this
case, the coherence time $T_2$ is determined by the bias current
noise in the junction, as in many superconducting qubits. We
estimated $T_\phi^{noise}$ by accounting for the fact that the
current noise may arise from the magnetic field fluctuations in
the junction. This bias current fluctuation is described
realistically by using the Gaussian colored noise with a long
correlation time.  Our estimated value of $T_\phi^{noise}$ for the
JVQ, which is obtained by using the experimental data from the
Nb-AlO$_x$-Nb FFO, is in the microsecond range because the
spectral density for fluctuating magnetic field is very low. The
value for $T_\phi$ is few orders of magnitude larger than that
measured for the quantronium, suggesting that the JVQ may also be
a good candidate for quantum computer. This surprisingly long
coherence time for the JVQ is due to the fact that the fluxon,
which behaves as a topologically stable quantum particle at
ultra-low temperature, couples very weakly to both internal and
external noise sources. The current estimate for $T_2$ may be
extended further if the magnetic field fluctuations, which are
considered as the dominant decoherence source in LJJ, can be
further reduced.

This work suggests possibility that a superonducting qubit with an
ultra-long coherence time may be realized by exploiting quantum
property of fluxon pinned in a double-well potential in LJJ.  This
work also provides insight into design and fabrication of the JVQ.
The current approach may be easily extended\cite{KM} to realize
multiple noninteracting qubits in quasi-one dimensional LJJ
stacks. Hence, it would be interesting to verify macroscopic
quantum coherence behavior by either spectroscopic measurement of
level splitting or by observation of Rabi oscillations in the JVQ.

\vskip 0.3cm

J.H.K. thanks K. S. Moon and H. J. Lee for useful discussions and
acknowledges financial support from NDEPSCoR through NSF Grant No.
EPS-0132289 and generosity and hospitality of Yonsei University
where part of this work was completed.  Also, the authors would
like to thank I. D. O'Bryant for assisting part of numerical
calculation.

\end{document}